\documentclass[aps,prb,twocolumn,superscriptaddress,
showpacs,preprintnumbers,amsmath,amssymb
,floatfix
]{revtex4}
\usepackage{graphicx}
\usepackage{dcolumn}
\usepackage{bm}
\usepackage{amsmath}
\usepackage{amssymb}
\usepackage{revsymb}
 
\begin{document}

\affiliation{
Department of Physics and Astronomy, Georgia State
University, Atlanta, Georgia 30303, USA}
\affiliation{
Department of Chemistry, MIT, Cambridge, MA 02139, USA}

\title{Toward Full Spatio-Temporal Control on the Nanoscale}
\affiliation{
Department of Physics and Astronomy, Georgia State
University, Atlanta, Georgia 30303, USA}

\author{Maxim Durach}
\affiliation{
Department of Physics and Astronomy, Georgia State
University, Atlanta, Georgia 30303, USA}
\author{Anastasia Rusina}
\affiliation{
Department of Physics and Astronomy, Georgia State
University, Atlanta, Georgia 30303, USA}
\author{Keith Nelson}
\affiliation{
Department of Chemistry, MIT, Cambridge, MA 02139, USA}
\author{Mark I. Stockman}
\affiliation{
Department of Physics and Astronomy, Georgia State
University, Atlanta, Georgia 30303, USA}
\email{mstockman@gsu.edu}
\homepage{http://www.phy-astr.gsu.edu/stockman}

\date{\today}

\begin{abstract}
We introduce an approach to implement full coherent control on 
nanometer length scales. It is
based on spatio-temporal modulation of the surface plasmon polariton
(SPP) fields at the thick edge of a nanowedge. The SPP wavepackets
propagating toward the sharp edge of this nanowedge are
compressed and adiabatically concentrated at a nanofocus, forming an ultrashort 
pulse of local fields. The one-dimensional spatial profile
and temporal waveform of this pulse are
completely coherently controlled.
\end{abstract}

\pacs{%
78.67.-n, %
%       Optical properties of nanoscale materials and structures
%
%78.45.+h
%       Optical properties, condensed-matter spectroscopy and other
%       interactions of radiation and particles with condensed matter
%       Stimulated emission (see also 42.55 Lasers)
%42.50.-p
%       Quantum optics (for lasers, see 42.55 and 42.60)
%
%78.20.Bh
%       Optical properties of bulk materials and thin films
%       Theory, models, and numerical simulation
%
%42.55.-f
%       Lasers
%
%68.37.Uv,%
%       Near-field scanning microscopy and spectroscopy
%
%78.47.+p,
%       Time-resolved optical spectroscopies and other ultrafast
%       optical measurements in condensed matter
%
%42.50.Md,
%       Optical transient phenomena: quantum beats, photon echo,
%       free-induction decay, dephasings and revivals, and
%       optical 
%	nutation, and self-induced transparency
%
%42.65.Sf,
%       Dynamics of nonlinear optical systems; optical instabilities,
%       optical chaos, and complexity, and optical spatio-temporal
%       dynamics
%
71.45.Gm,
%       Electron structure:
%               Exchange, correlation, dielectric and magnetic
%       functions, plasmons
%
42.65.Re,
%       Ultrafast processes; optical pulse generation and pulse compression
%
%05.45.+b,
%       Statistical physics and thermodynamics:
%       Theory and models of chaotic systems
%
%61.43.Hv,
%       Condensed matter, Structure of solids and liquids:
%       Disordered solids. Fractals, Macroscopic
%       Aggregates (including diffusion-limited aggregates).
%
%
%05.40.+j,
%       Statistical physics and thermodynamics:
%       Fluctuation phenomena, random processes, and Brownian
%       motion
%73.20 Fz,
%       Weak or Anderson localization. Surfaces, interfaces,
%       thin films and low-dimensional structures
%
73.20.Mf%
%       Collective excitations (including excitons,
%       polarons, plasmons and other charge-density excitations)
%
%85.35-p,
%       Nanoelectronic devices
%
%61.43.Hv,
%       Condensed matter, Structure of solids and liquids:
%       Disordered solids. Fractals, Macroscopic
%       Aggregates (including diffusion-limited aggregates).
%
%
%05.40.+j,
%       Statistical physics and thermodynamics:
%       Fluctuation phenomena, random processes, and Brownian
%       motion
}

\maketitle

Two novel areas of optics have recently attracted a great deal of
attention: nanooptics and ultrafast optics. One of the most rapidly
developing directions in ultrafast optics is 
quantum control, in which coherent superpositions of quantum states are
created by excitation radiation to control the quantum dynamics and
outcomes
\cite{Shapiro_Brumer_Phys_Rep_415_195_2006_Quantum_Control_of_Dynamics,
%Nguyen_Dey_Shapiro_Brumer_JPCA_108_7878_2004,
Rabitz_et_al_Science_288_824_2000_Quantum_Control,
%Geremia_Rabitz_PRL_89_263902_2002_Optimal_Hamiltonian_Identification,
%Krausz_et_al_Nature_427_817_2004_Atomic_Transient_Recorder,
Apolonski_Krausz_et_al_Phys_Rev_Lett_92_073902_2004_Phase_Sensitive_Photoemission,
%Niikura_Villeneuve_Corkum_Phys_Rev_Lett_94_083003_2005,
Corkum_et_al_PRL_95_203003_2005_Controling_as_Double_Ionization}. Of
special interest are coherently controlled ultrafast phenomena on the
nanoscale where the phase of the excitation waveform along with its
polarization provides a
functional degree of freedom to control nanoscale distribution of energy
\cite{Stockman:2002_PRL_control,
Phys_Rev_B_69_054202_2004_Stockman_Bergman_Kobayashi_Coherent_Control,
Stockman_Hewageegana_Nano_Lett_2005_V_Shape_Coherent_Control,
Kubo_Onda_Petek_Sun_Jung_Kim_Nano_Lett_2005_Two_Pulse_Coherent_Control,
%Brixner_et_al_PRL_92_208301_2004_Polarization_Quantum_Control,
Sukharev_Seideman_Nano_Lett_2006_Phase_Polarization_Control,
Aeschlimann_Bauer_Bayer_Brixner_et_al_Nature_446_301_2007_Nanooptical_Adaptive_Control}. 
Spatiotemporal pulse shaping
permits one to generate dynamically predefined waveforms modulated
both in frequency and in space to focus
ultrafast pulses in the required microscopic spatial and 
femtosecond temporal domains
\cite{Wefers_Nelson_Opt_Lett_23_2032_1993_Spatiotemporal_Pulse_Shaping,
Feurer_Vaughan_Nelson:2003_Science}.

In this Letter, we propose and theoretically develop a method of full
coherent control on the nanoscale where a spatiotemporally modulated pulse
is launched in a graded nanostructured system. Its propagation and
adiabatic concentration provide a possibility to focus the optical
energy in {\it nanoscale} spatial and
femtosecond temporal regions. The idea of adiabatic concentration 
\cite{Phys_Rev_Lett_93_2004_Tapered_Plasmonic_Waveguides,
Stockman_Plasmonics_2004_Adiabatic_Concentration_SPIE} 
(see also Ref.\
\onlinecite{Babajanyan_Margaryan_Nerkararyan_JAP_2000_SPP_on_Cone}) 
is based on adiabatic following by the propagating surface
plasmon-polariton (SPP) wave of a plasmonic waveguide, where  
the phase and group velocities decrease
toward a limit at which the propagating SPP wave is adiabatically
transformed into a standing surface plasmon (SP) mode.
This effect has 
been further developed theoretically
\cite{Maier_Andrews_Martin-Moreno_Garcia-Vidal_PRL_2006_THz_SPPs_in_Periodically_Corrugated_Wires,
Gramotnev_JAP_2005_Adiabatic_SPP_Nanofocusing} and observed
experimentally.
\cite{Verhagen_Kuipers_Polman_Nano_Lett_2006_Tapered_Plasmonic_Waveguide}

%--------------------------------------------------------------------
\begin{figure}
\centering
\includegraphics[width=.45\textwidth]
{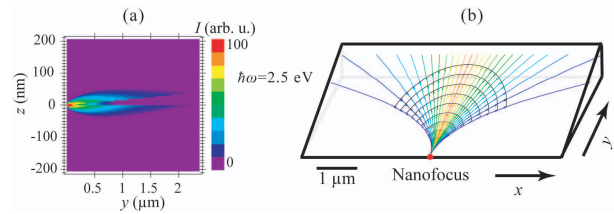}
\caption{\label{Concentration_and_Trajectories.eps}
(a) Illustration of adiabatic concentration of energy on the wedge.
The distribution of local field intensity $I$ in the normal plane of
propagation of SPPs (the $yz$ plane). The intensity in relative units 
is color coded with
the color scale bar shown to the right.
(b) Trajectories of SPP rays propagating from the thick to sharp edge of
the wedge. The initial coordinate is coded with color. The black curves
indicate lines of equal phase (SPP wave fronts).
%The SPP frequency is
%as shown, $\hbar\omega=2.5~\mathrm{eV}$.
 }
\end{figure}
%--------------------------------------------------------------------

To illustrate the idea of this full coherent control, consider first the
adiabatic concentration of a plane SPP wave propagating along a nanowedge of
silver \cite{Johnson:1972_Silver}, 
as shown in Fig.\ \ref{Concentration_and_Trajectories.eps}(a);
the theory is based on the Wentzel-Kramers-Brillouin (WKB) or
quasiclassical approximation, also called the eikonal approximation in
optics \cite{Landau_Lifshitz_Electrodynamics_Continuous:1984} as
suggested in  Refs.\ 
\onlinecite{Phys_Rev_Lett_93_2004_Tapered_Plasmonic_Waveguides,
Stockman_Plasmonics_2004_Adiabatic_Concentration_SPIE}. The propagation
velocity of the SPP along such a nanowedge is asymptotically
proportional to its thickness. Thus when a SPP approaches the sharp
edge, it slows down and asymptotically (in the ideal limit of zero 
thickness at the apex) stops while the local fields are
increased and nano-concentrated. 

Now consider a family of SPP rays (WKB trajectories) propagating from
the thick side of a nanowedge, as shown in Fig.\
\ref{Concentration_and_Trajectories.eps}(b). One possibility to launch
such SPPs is to have nanoscale inhomogeneities (nanoparticles or
nanoholes) at the thick edge of the wedge. Each of the optical pulses
from the spatiotemporal shaper is focused on the corresponding
nanoparticle, scattering from it and generating an SPP wave. This
scattering is necessary to impart large transverse momenta on SPPs,
which is required for their efficient transverse focusing. These SPP
waves propagate toward the sharp edge, adiabatically slow down which
increases the field amplitudes, and constructively interfere 
as they converge at the nanofocus. 

The phases of the SPPs rays (i.e., the corresponding wave
fronts) are defined by the spatiotemporal modulator in such a way that
the rays converge to a reconfigurably chosen point at the sharp edge of the
nanowedge where they acquire equal phases. When the SPPs propagate along
the rays, adiabatic concentration takes place because the SPP wavelength
tends to zero proportionally to the thickness of the wedge. This allows one
to focus optical energy at a {\it predefined} nanofocus at the sharp
edge. The temporal structure of the generated SPPs can be chosen in such
a way that at the nanofocus the local fields form an ultrashort pulse.
Using spatiotemporal modulation of the excitation field at the thick
edge, one can arbitrarily move the nanofocus along the sharp edge. A
superposition of such fields can render arbitrary spatiotemporal
modulation on the sharp edge, enabling one to exert full control over
nanoscale fields in space and time.

Turning to the theory, 
consider a nanofilm of metal in an $xy$ plane whose thickness
$d$ in the $z$ direction is adiabatically changing with the
coordinate-vector ${\bm \rho}=(x,y)$ in the plane of the nanofilm. Let
$\varepsilon_m=\varepsilon_m(\omega)$ be the dielectric permittivity of
this metal nanofilm, and $\varepsilon_d$ be the permittivity of the
embedding dielectric. Because of the symmetry of the system, there are
odd and even (in the normal electric field) SPPs. It is the odd SPP that
is a slow-propagating, controllable mode. The dispersion relation for
this mode defining its effective index $n({\bm\rho})$ is
\begin{equation}
\tanh\left(\frac{1}{2} k_0 d({\bm\rho})\sqrt{n({\bm\rho})^2-\varepsilon_m}\right)=
-\frac{\varepsilon_d\sqrt{n({\bm\rho})^2-\varepsilon_m}}
{\varepsilon_m\sqrt{n({\bm\rho})^2-\varepsilon_d}}~,
\label{disrel}
\end{equation}
where $k_0=\omega/c$ is the radiation wave vector in vacuum.

Let $\bm\tau$ be a unit tangential vector to the SPP trajectory (ray).
It obeys a conventional equation of ray optics 
\cite{Landau_Lifshitz_Electrodynamics_Continuous:1984}
\begin{equation}
n \frac{d {\bm \tau}}{d l}=
\nabla n - {\bm \tau} \left({\bm \tau} \nabla n \right)~,
\label{rayeqn}
\end{equation}
where $l$ is the length along the ray and
$\nabla=\partial/\partial{\bm\rho}$. 

Now let us consider a nanofilm shaped as a nanowedge as in Fig.\ 
\ref{Concentration_and_Trajectories.eps}(b). In such a case, 
$n=n(y)$, and these 
trajectory equations simplify as
\begin{equation}
n\frac{d\tau_y}{d l}=\tau_x^2\frac{\partial n}{\partial y}~,~~~
n\frac{d\tau_x}{d l}=-\tau_x\tau_y\frac{\partial n}{\partial y}~.
\end{equation}
From these, it follows that $n_x\equiv\tau_x n=\mathrm{const}$.
The SPP wave vector, related to its momentum, is
$\mathbf k({\bm\rho})=k_0 n({\bm\rho}){\bm \tau}$; this is the
conservation of $k_x$ (the transverse momentum). This
allows one to obtain a closed solution for the ray. The tangent
equation for the ray is
$d x/d y=\tau_x/\tau_y$, where $\tau_y=\sqrt{1-n_x^2/n^2}$. From this,
we get an explicit SPP trajectory (ray) equation as
\begin{equation}
x-x_0=\int_{y_0}^y \left(\frac{n(y^\prime)^2}{n_x^2}-1\right)^{-1/2}
d y^\prime~,
\label{trajectory_eq}
\end{equation}
where $\bm\rho_0=(x_0,y_0)$ is the focal point where rays with any $n_x$
converge. To find the trajectories, as $n(y)$ we use the
real part of effective index (\ref{disrel}), as WKB suggests.

When the local thickness of the wedge is
subwavelength ($k_0 d\ll 1$), the form of these trajectories can be
found analytically. Under these conditions, 
dispersion relation (\ref{disrel}) has an asymptotic solution
\begin{equation}
n=\frac{n_a}{k_0 d}~,~~~
n_a=\ln{\frac{\varepsilon_m-\varepsilon_d}
{\varepsilon_m+\varepsilon_d}}~.
\label{asymptotic_disrel}
\end{equation}
Substituting this into Eq.\ (\ref{trajectory_eq}), we obtain explicit equations
of trajectories,
\begin{equation}
\left(x-x_0-\sqrt{\frac{\bar n_a^2}{n_x^2}-y_0^2}\right)^2+y^2=
\frac{\bar n_a^2}{n_x^2}~,
\end{equation}
where $\bar n_a=n_a/(k_0 \tan\theta)$, and $\tan\theta$ is the slope of the
wedge. Thus, each SPP ray is a segment of a circle whose center is at
a point given by $x=x_0+\sqrt{(\bar n_a/n_x)^2-y_0^2}$ and $y=0$. 
This analytical result is in agreement with 
Fig.\ \ref{Concentration_and_Trajectories.eps} (b). If
the nanofocus is at the sharp edge, i.e., $y_0=0$, then these circles do
not intersect but touch and are tangent to each other at the nanofocus point.

As an example we consider a silver \cite{Johnson:1972_Silver} nanowedge
illustrated in Fig.\ \ref{Concentration_and_Trajectories.eps} (b) whose
maximum thickness is $d_m=30$ nm and whose length (in the $y$ direction)
is $L=5~\mu\mathrm{m}$. Trajectories calculated from Eq.\
(\ref{trajectory_eq}) for $\hbar\omega=2.5$ eV
are shown by lines (color used only to guide eye); 
the nanofocus is indicated by a bold red dot. The different
trajectories correspond to different values of $n_x$ in the range $0\le
n_x\le n(L)$. In contrast to focusing by a conventional lens, the SPP
rays are progressively bent toward the wedge slope direction. 

%--------------------------------------------------------------------
\begin{figure}
\centering
\includegraphics[width=.45\textwidth]
{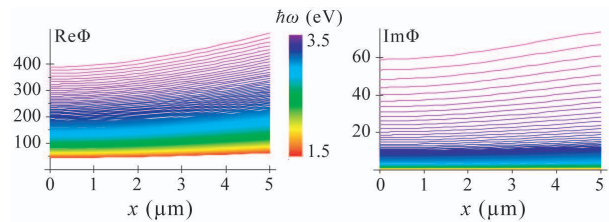}
\caption{\label{Eikonal.eps}
(a) Phase (real part of eikonal $\Phi$)
acquired by a SPP ray propagating between a point with
coordinate $x$ on the thick edge and the nanofocus, displayed as a
function of $x$. The rays differ by frequencies that are color coded by
the vertical bar.
(b) The same as (a) but for extinction of the
ray ($\mathrm{Im}\,\Phi$).
}
\end{figure}
%--------------------------------------------------------------------

The eikonal is found as an integral along the ray \begin{equation}
\Phi({\bm\rho})=\int^{\bm\rho}_{{\bm\rho}_0} \mathbf n({\bm\rho})d{\bm\rho}~.
\label{eikonal} \end{equation} Consider rays emitted from the nanofocus
[Fig.\ \ref{Concentration_and_Trajectories.eps} (b)]. Computed from this
equation for frequencies in the visible range, the phases of the SPPs at
the thick edge of the wedge (for $y=L$) are shown in Fig.\
\ref{Eikonal.eps} (a) as functions of the coordinate $x$ along the thick
edge. The colors of the rays correspond to the visual perception of the
ray frequencies. 
%Note that the central ray (at each frequency)
%propagating along the $y$ direction (gradient of the wedge slope)
%corrresponds to $x=0$. 
The gained phase dramatically increases toward the blue
spectral region, exhibiting a strong dispersion. The extinction
for most of the frequencies except for the blue edge, displayed
in Fig.\ \ref{Eikonal.eps} (b), is not high.

Now consider the evolution of the field intensity along a SPP ray. For
certainty, let SPPs propagate along the corresponding rays from the thick
edge of the wedge toward the nanofocus as shown in Fig.\ 
\ref{Concentration_and_Trajectories.eps} (b). In the process of such
propagation, there will be concentration of the SPP energy in all three
directions (3d nanofocusing). This phenomenon differs dramatically from
what occurs in conventional photonic ray optics. 

To describe this nanofocusing, it is convenient to introduce an
orthogonal system of ray coordinates whose unit vectors are
$\bm\tau$ (along the ray), $\bm\eta=(-\tau_y,\tau_x)$ (at the surface normal
to the ray), and $\mathbf e_z$ (normal to the surface).
The concentration
along the ray (in the $\bm\tau$ direction) occurs because the group
velocity $v_g=\left[\partial(k_0 n)/\partial\omega\right]^{-1}$ 
of SPP asymptotically tends to zero (for the
antisymmetric mode) for $k_0 d\to 0$
as $v_g= v_{0g} d$ where $v_{0g}=\mathrm{const}$.
\cite{Phys_Rev_Lett_93_2004_Tapered_Plasmonic_Waveguides,
Stockman_Plasmonics_2004_Adiabatic_Concentration_SPIE} This contributes
a factor $A_\parallel=1/\sqrt{v_g(d)}$ to the amplitude of an SPP wave. 

The compression of a SPP wave in the $\mathbf e_z$
(vertical) direction is given by a factor of $A_z=
\left(\int_{-\infty}^\infty W \mathrm d z\right)^{-1/2}$, where $W$ is
the energy density of the mode. Substituting a standard expression 
\cite{Landau_Lifshitz_Electrodynamics_Continuous:1984}
for $W$, one obtains explicitly
\begin{eqnarray}
&A_z=\Bigg(\frac{1}{8\pi}\exp\left(\mathrm{Re}\,\kappa_d d\right)\Big\{&
\nonumber\\
&\displaystyle\frac{\sinh\left(\mathrm{Re}\,\kappa_m d\right)}
{\mathrm{Re}\,\kappa_m \left|\sinh\left(\kappa_m d/2\right)\right|^2}
\left[1+\frac{d(\omega\mathrm{Re}\,\varepsilon_m)}{d\omega}
\frac{|n|^2+|\kappa_m|^2}{|\varepsilon_m|^2}\right]-&
\nonumber\\
&\displaystyle\frac{\sin\left(\mathrm{Im}\,\kappa_m d\right)}
{\mathrm{Im}\,\kappa_m \left|\sinh\left(\kappa_m d/2\right)\right|^2}
\left[1+\frac{d(\omega\mathrm{Re}\,\varepsilon_m)}{d\omega}
\frac{|n|^2-|\kappa_m|^2}{|\varepsilon_m|^2}\right]+&
\nonumber\\
&\displaystyle\frac{2}{\mathrm{Re}\,\kappa_d}
\left[1+\frac{|n|^2+|\kappa_d|^2}{\varepsilon_d}\right]\Big\}\Bigg)^{-1/2},&
\end{eqnarray}
where $\kappa_m=k_0\sqrt{n-\varepsilon_m}$ and
$\kappa_d=k_0\sqrt{n-\varepsilon_d}$.

To obtain the compression factor $A_\perp$ for the $\bm\eta$ direction),
we consider conservation of energy along the beam of rays corresponding
to slightly different values of $n_x$. Dividing this constant energy
flux by the thickness of this beam in the $\bm\eta$ direction, we arrive
at 
\begin{equation}
A_\perp=\left\{\left(1-\frac{n_x^2}{n^2}\right)^{1/2}%
\int_{y_0}^y\frac{1}{n(y^\prime)}%
\left[1-\frac{n_x^2}{n(y^\prime)^2}\right]^{-3/2}%
\mathrm d y^\prime \right\}^{-1/2}.
\end{equation}
The ray amplitude thus contains the total factor which describes the
3d adiabatic compression: $A=A_\parallel A_\perp A_z$.

%--------------------------------------------------------------------
\begin{figure}
\centering
\includegraphics[width=.45\textwidth]
{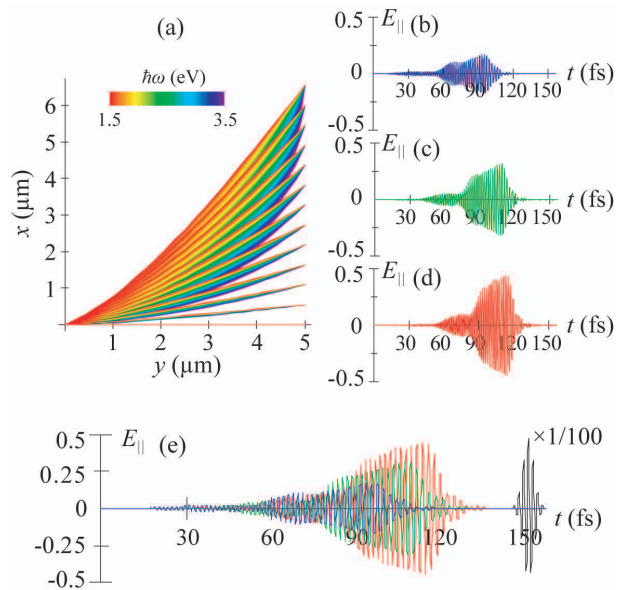}
\caption{\label{Trajectories_and_Waveforms.eps}
(a) Trajectories (rays) of SPP packets propagating from the thick edge to the
nanofocus displayed
in the $xy$ plane of the wedge. The frequencies of the
individual rays in a packet are indicated by color as coded by the 
bar at the top. (b)-(d) Spatiotemporal
modulation of the excitation pulses at the thick edge of the wedge
required for nanofocusing. The temporal dependencies (waveforms) of the
electric field for the phase-modulated pulses for three points at the
thick edge boundary: two extreme points and one at the center, as
indicated, aligned with the corresponding $x$ points at panel (a). 
(e) The three excitation pulses of panels (b)-(d) (as shown by their
colors), superimposed to
elucidate the phase shifts, delays, and shape changes between these pulses.
The resulting ultrashort pulse at the nanofocus is shown by the 
black line. The scale of the
electric fields is arbitrary but consistent throughout the figure.
}
\end{figure}
%--------------------------------------------------------------------

Now consider the problem of coherent control.
The goal is to excite a spatiotemporal waveform at the thick edge of
the wedge in such a way that the propagating SPP rays converge at
an arbitrary nanofocus at the sharp edge where an ultrashort pulse is formed. 
To solve this problem, we use the idea of
back-propagation or time-reversal.
\cite{%Fink_et_al_PRL_92_193904_2004_Time_Reversal_of_EM_Waves,
%Fink_et_al_APL_88_15401_2006_Wideband_Microwave_Time_Reversal,
Lerosey_de_Rosny_Tourin_Fink_Science_315_1120_2007_Microwave_Time%
_Reversal_Subwavelength_Focusing}
We generate rays at the nanofocus as an ultrashort pulse
containing just several oscillations of the light field. Propagating these
rays, we find amplitudes and phases of
the fields at the thick edge at each frequency as given by the eikonal
$\Phi(\bm\rho)$. Then we complex conjugate the amplitudes of frequency
components, which corresponds to the time reversal. We also multiply 
these amplitudes by $\exp(2 \mathrm{Im}\,\Phi)$ which pre-compensates
for the losses. This provides the required phase and
amplitude modulation at the thick edge of the wedge. 

We show an example of such calculations in Fig.\
\ref{Trajectories_and_Waveforms.eps}. Panel (a) displays the
trajectories of SPPs calculated according to Eq.\ (\ref{trajectory_eq}).
The trajectories for different frequencies are displayed by colors
corresponding to their visual perception. There is a very
significant spectral dispersion: trajectories with higher frequencies
are much more curved. The spatial-frequency modulation that we have
found succeeds in bringing all
these rays (with different frequencies and emitted at different $x$
points) to the same nanofocus at the sharp edge.

The required waveforms at different $x$ points of the thick edge of the
wedge are shown in Fig.\ \ref{Trajectories_and_Waveforms.eps} (b)-(d)
where the corresponding longitudinal electric fields are shown. 
The waves emitted at large $x$, i.e., at points more distant from the
nanofocus, should be emitted significantly earlier to pre-compensate for the
longer propagation times. They should also have different amplitudes
due to the differences in $A$.
Finally, there is clearly a negative chirp (gradual decrease of
frequency with time).
This is due to the fact that the higher frequency
components propagate more slowly and therefore must be emitted earlier to
form a coherent ultrashort pulse at the nanofocus. 

In Fig.\ \ref{Trajectories_and_Waveforms.eps} (e) we display together
all three of the representative waveforms at the thick edge to
demonstrate their relative amplitudes and positions in time. 
The pulse at the extreme point in $x$ (shown by blue) has the longest
way to propagate and therefore is the most advanced in time. 
The pulse in the middle point
(shown by green) is intermediate, and the pulse at the center
($x=0$, shown by red) is last.
One can notice also a counterintuitive
feature: the waves propagating over longer trajectories are
smaller in amplitude though one may expect the opposite to
compensate for the larger losses. The explanation of this fact is that
the losses are actually insignificant for the frequencies present in
these waveforms. What determines the relative magnitudes of these
waveforms is the coefficient $A_\perp$ of the transverse concentration,
which is much higher for the peripheral trajectories than for the
central ones.

Figure \ref{Trajectories_and_Waveforms.eps} (e) also shows the resulting
ultrashort pulse in the nanofocus. This is a transform-limited,
Gaussian pulse. The propagation along the rays completely
compensates the initial phase and amplitude modulation, exactly as
intended. As a result, the corresponding electric field of the waveform
is increased by a factor of $100$. Taking the other component
of the electric field and the magnetic field into account, the
corresponding increase of the energy density is by a factor $\sim10^4$
with respect to that of the SPPs at the thick edge.

Consider the efficiency of the energy transfer to the nanoscale.
This is primarily determined by the cross section
$\sigma_{_{SPP}}$ for scattering of photons into SPPs. For instance, for
a metal sphere of radius $R$ at the surface of the wedge, one can obtain
an estimate $\sigma_{_{SPP}}\sim R^6/(d_m^3\lambdabar)$, where
$\lambdabar$ is the reduced photon wavelength. Setting $R\sim d_m$, we
estimate $\sigma_{_{SPP}}\sim 3 ~\mathrm{nm^2}$. Assuming optical
focusing into a spot of $\sim 300$ nm radius, this yields the energy
efficiency of conversion to the nanoscale of $\sim 10^{-3}$. Taking
into account the adiabatic concentration of energy by a factor of
$10^4$, the optical field intensity at the nanofocus is
enhanced by one order of magnitude with respect to that of the incoming
optical wave.

The criterion of applicability of the WKB approximation is $\partial
k^{-1}/\partial y\ll 1$. Substituting $k=k_0 n$ and Eq.\
(\ref{asymptotic_disrel}), we obtain a condition $d_m/(n_a L)<<1$. This
condition is satisfied everywhere including the nanofocus since $n_a\sim
1$ and $d_m\ll L$ for adiabatic grading. The minimum possible size
of the wavepacket at the nanofocus in the direction of propagation,
$\Delta x$, is limited by the local SPP wavelength: $\Delta x\sim
2\pi/k\approx 2\pi d_f/n_a$, where $d_f$ is the wedge thickness at the
nanofocus. The minimum transverse size $a$ (waist) of the SPP beam at
the nanofocus can be calculated as the radius of the first Fresnel
zone: $a=\pi/k_x\ge \pi/(k_0 n_x)$. Because $n_x$ is constant along a
trajectory, one can substitute its value at the thick edge (the launch
site), where from Eq.\ (\ref{asymptotic_disrel}) we obtain $n_x\approx
n=n_a/d_m$. This results in $a\approx \pi d_m/n_a$; thus $a$ is on
order of the maximum thickness of the wedge, which is assumed also to be on
the nanoscale. 

%Another important characteristic of the nanofocus is the
%confocal parameter, i.e., the length of the focal waist in the
%longitudinal direction $b=a L/d_m\approx 2\pi L/n_a$. Thus the waist
%diameter and the minimum wave packet length at the nanofocus are of the
%same size, on order of the wedge maximum thickness: 
%$\Delta_x\approx a\approx 2\pi d_m/n_a$. The confocal parameter $b$ is much
%larger, on order of the length of the wedge. 

To briefly conclude, we have proposed and theoretically investigated an
approach to full coherent control of spatiotemporal
energy localization on the nanoscale. From the thick edge of a plasmonic
metal nanowedge, SPPs are launched, whose phases and amplitudes are
independently modulated for each constituent frequency of the spectrum
and at each spatial point of the excitation. This pre-modulates the
departing SPP wave packets in such a way that they reach the required
point at the sharp edge of the nanowedge in phase, with equal amplitudes
forming a nanofocus where an ultrashort pulse with required temporal shape is
generated. This system constitutes a ``nanoplasmonic portal'' connecting
the incident light field, whose features are shaped on the microscale, 
with the required point or features at the nanoscale.

This work was supported by grants from the Chemical Sciences, Biosciences and
Geosciences Division of the Office of Basic Energy Sciences, Office of
Science, U.S. Department of Energy, a grant CHE-0507147 from NSF, 
and a grant from the US-Israel BSF.


\begin{thebibliography}{21}
\expandafter\ifx\csname natexlab\endcsname\relax\def\natexlab#1{#1}\fi
\expandafter\ifx\csname bibnamefont\endcsname\relax
  \def\bibnamefont#1{#1}\fi
\expandafter\ifx\csname bibfnamefont\endcsname\relax
  \def\bibfnamefont#1{#1}\fi
\expandafter\ifx\csname citenamefont\endcsname\relax
  \def\citenamefont#1{#1}\fi
\expandafter\ifx\csname url\endcsname\relax
  \def\url#1{\texttt{#1}}\fi
\expandafter\ifx\csname urlprefix\endcsname\relax\def\urlprefix{URL }\fi
\providecommand{\bibinfo}[2]{#2}
\providecommand{\eprint}[2][]{\url{#2}}

\bibitem[{\citenamefont{Shapiro and
  Brumer}(2006)}]{Shapiro_Brumer_Phys_Rep_415_195_2006_Quantum_Control_of_Dyna%
mics}
\bibinfo{author}{\bibfnamefont{M.}~\bibnamefont{Shapiro}} \bibnamefont{and}
  \bibinfo{author}{\bibfnamefont{P.}~\bibnamefont{Brumer}},
  \bibinfo{journal}{Physics Reports} \textbf{\bibinfo{volume}{425}},
  \bibinfo{pages}{195} (\bibinfo{year}{2006}).

\bibitem[{\citenamefont{Rabitz et~al.}(2000)\citenamefont{Rabitz,
  de~Vivie-Riedle, Motzkus, and
  Kompa}}]{Rabitz_et_al_Science_288_824_2000_Quantum_Control}
\bibinfo{author}{\bibfnamefont{H.}~\bibnamefont{Rabitz}},
  \bibinfo{author}{\bibfnamefont{R.}~\bibnamefont{de~Vivie-Riedle}},
  \bibinfo{author}{\bibfnamefont{M.}~\bibnamefont{Motzkus}}, \bibnamefont{and}
  \bibinfo{author}{\bibfnamefont{K.}~\bibnamefont{Kompa}},
  \bibinfo{journal}{Science} \textbf{\bibinfo{volume}{288}},
  \bibinfo{pages}{824} (\bibinfo{year}{2000}).

\bibitem[{\citenamefont{Apolonski et~al.}(2004)\citenamefont{Apolonski, Dombi,
  Paulus, Kakehata, Holzwarth, Udem, Lemell, Torizuka, Burgdoerfer, Hansch
  et~al.}}]{Apolonski_Krausz_et_al_Phys_Rev_Lett_92_073902_2004_Phase_Sensitiv%
e_Photoemission}
\bibinfo{author}{\bibfnamefont{A.}~\bibnamefont{Apolonski}},
  \bibinfo{author}{\bibfnamefont{P.}~\bibnamefont{Dombi}},
  \bibinfo{author}{\bibfnamefont{G.~G.} \bibnamefont{Paulus}},
  \bibinfo{author}{\bibfnamefont{M.}~\bibnamefont{Kakehata}},
  \bibinfo{author}{\bibfnamefont{R.}~\bibnamefont{Holzwarth}},
  \bibinfo{author}{\bibfnamefont{T.}~\bibnamefont{Udem}},
  \bibinfo{author}{\bibfnamefont{C.}~\bibnamefont{Lemell}},
  \bibinfo{author}{\bibfnamefont{K.}~\bibnamefont{Torizuka}},
  \bibinfo{author}{\bibfnamefont{J.}~\bibnamefont{Burgdoerfer}},
  \bibinfo{author}{\bibfnamefont{T.~W.} \bibnamefont{Hansch}},
  \bibnamefont{et~al.}, \bibinfo{journal}{Phys. Rev. Lett.}
  \textbf{\bibinfo{volume}{92}}, \bibinfo{pages}{073902}
  (\bibinfo{year}{2004}).

\bibitem[{\citenamefont{Zeidler et~al.}(2005)\citenamefont{Zeidler, Staudte,
  Bardon, Villeneuve, Dorner, and
  Corkum}}]{Corkum_et_al_PRL_95_203003_2005_Controling_as_Double_Ionization}
\bibinfo{author}{\bibfnamefont{D.}~\bibnamefont{Zeidler}},
  \bibinfo{author}{\bibfnamefont{A.}~\bibnamefont{Staudte}},
  \bibinfo{author}{\bibfnamefont{A.~B.} \bibnamefont{Bardon}},
  \bibinfo{author}{\bibfnamefont{D.~M.} \bibnamefont{Villeneuve}},
  \bibinfo{author}{\bibfnamefont{R.}~\bibnamefont{Dorner}}, \bibnamefont{and}
  \bibinfo{author}{\bibfnamefont{P.~B.} \bibnamefont{Corkum}},
  \bibinfo{journal}{Phys. Rev. Lett.} \textbf{\bibinfo{volume}{95}},
  \bibinfo{pages}{203003} (\bibinfo{year}{2005}).

\bibitem[{\citenamefont{Stockman et~al.}(2002)\citenamefont{Stockman, Faleev,
  and Bergman}}]{Stockman:2002_PRL_control}
\bibinfo{author}{\bibfnamefont{M.~I.} \bibnamefont{Stockman}},
  \bibinfo{author}{\bibfnamefont{S.~V.} \bibnamefont{Faleev}},
  \bibnamefont{and} \bibinfo{author}{\bibfnamefont{D.~J.}
  \bibnamefont{Bergman}}, \bibinfo{journal}{Phys. Rev. Lett.}
  \textbf{\bibinfo{volume}{88}}, \bibinfo{pages}{067402}
  (\bibinfo{year}{2002}).

\bibitem[{\citenamefont{Stockman et~al.}(2004)\citenamefont{Stockman, Bergman,
  and
  Kobayashi}}]{Phys_Rev_B_69_054202_2004_Stockman_Bergman_Kobayashi_Coherent_C%
ontrol}
\bibinfo{author}{\bibfnamefont{M.~I.} \bibnamefont{Stockman}},
  \bibinfo{author}{\bibfnamefont{D.~J.} \bibnamefont{Bergman}},
  \bibnamefont{and}
  \bibinfo{author}{\bibfnamefont{T.}~\bibnamefont{Kobayashi}},
  \bibinfo{journal}{Phys. Rev. B} \textbf{\bibinfo{volume}{69}},
  \bibinfo{pages}{054202} (\bibinfo{year}{2004}).

\bibitem[{\citenamefont{Stockman and
  Hewageegana}(2005)}]{Stockman_Hewageegana_Nano_Lett_2005_V_Shape_Coherent_Co%
ntrol}
\bibinfo{author}{\bibfnamefont{M.~I.} \bibnamefont{Stockman}} \bibnamefont{and}
  \bibinfo{author}{\bibfnamefont{P.}~\bibnamefont{Hewageegana}},
  \bibinfo{journal}{Nano Lett.} \textbf{\bibinfo{volume}{5}},
  \bibinfo{pages}{2325} (\bibinfo{year}{2005}).

\bibitem[{\citenamefont{Kubo et~al.}(2005)\citenamefont{Kubo, Onda, Petek, Sun,
  Jung, and
  Kim}}]{Kubo_Onda_Petek_Sun_Jung_Kim_Nano_Lett_2005_Two_Pulse_Coherent_Contro%
l}
\bibinfo{author}{\bibfnamefont{A.}~\bibnamefont{Kubo}},
  \bibinfo{author}{\bibfnamefont{K.}~\bibnamefont{Onda}},
  \bibinfo{author}{\bibfnamefont{H.}~\bibnamefont{Petek}},
  \bibinfo{author}{\bibfnamefont{Z.}~\bibnamefont{Sun}},
  \bibinfo{author}{\bibfnamefont{Y.~S.} \bibnamefont{Jung}}, \bibnamefont{and}
  \bibinfo{author}{\bibfnamefont{H.~K.} \bibnamefont{Kim}},
  \bibinfo{journal}{Nano Lett.} \textbf{\bibinfo{volume}{5}},
  \bibinfo{pages}{1123} (\bibinfo{year}{2005}).

\bibitem[{\citenamefont{Sukharev and
  Seideman}(2006)}]{Sukharev_Seideman_Nano_Lett_2006_Phase_Polarization_Contro%
l}
\bibinfo{author}{\bibfnamefont{M.}~\bibnamefont{Sukharev}} \bibnamefont{and}
  \bibinfo{author}{\bibfnamefont{T.}~\bibnamefont{Seideman}},
  \bibinfo{journal}{Nano Lett.} \textbf{\bibinfo{volume}{6}},
  \bibinfo{pages}{715} (\bibinfo{year}{2006}).

\bibitem[{\citenamefont{Aeschlimann et~al.}(2007)\citenamefont{Aeschlimann,
  Bauer, Bayer, Brixner, d.~Abajo, Pfeiffer, Rohmer, Spindler, and
  Steeb}}]{Aeschlimann_Bauer_Bayer_Brixner_et_al_Nature_446_301_2007_Nanooptic%
al_Adaptive_Control}
\bibinfo{author}{\bibfnamefont{M.}~\bibnamefont{Aeschlimann}},
  \bibinfo{author}{\bibfnamefont{M.}~\bibnamefont{Bauer}},
  \bibinfo{author}{\bibfnamefont{D.}~\bibnamefont{Bayer}},
  \bibinfo{author}{\bibfnamefont{T.}~\bibnamefont{Brixner}},
  \bibinfo{author}{\bibfnamefont{F.~J.~G.} \bibnamefont{d.~Abajo}},
  \bibinfo{author}{\bibfnamefont{W.}~\bibnamefont{Pfeiffer}},
  \bibinfo{author}{\bibfnamefont{M.}~\bibnamefont{Rohmer}},
  \bibinfo{author}{\bibfnamefont{C.}~\bibnamefont{Spindler}}, \bibnamefont{and}
  \bibinfo{author}{\bibfnamefont{F.}~\bibnamefont{Steeb}},
  \bibinfo{journal}{Nature} \textbf{\bibinfo{volume}{446}},
  \bibinfo{pages}{301} (\bibinfo{year}{2007}).

\bibitem[{\citenamefont{Wefers and
  Nelson}(1993)}]{Wefers_Nelson_Opt_Lett_23_2032_1993_Spatiotemporal_Pulse_Sha%
ping}
\bibinfo{author}{\bibfnamefont{M.~M.} \bibnamefont{Wefers}} \bibnamefont{and}
  \bibinfo{author}{\bibfnamefont{K.~A.} \bibnamefont{Nelson}},
  \bibinfo{journal}{Opt. Lett.} \textbf{\bibinfo{volume}{18}},
  \bibinfo{pages}{2032} (\bibinfo{year}{1993}).

\bibitem[{\citenamefont{Feurer et~al.}(2003)\citenamefont{Feurer, Vaughan, and
  Nelson}}]{Feurer_Vaughan_Nelson:2003_Science}
\bibinfo{author}{\bibfnamefont{T.}~\bibnamefont{Feurer}},
  \bibinfo{author}{\bibfnamefont{J.~C.} \bibnamefont{Vaughan}},
  \bibnamefont{and} \bibinfo{author}{\bibfnamefont{K.~A.}
  \bibnamefont{Nelson}}, \bibinfo{journal}{Science}
  \textbf{\bibinfo{volume}{299}}, \bibinfo{pages}{374} (\bibinfo{year}{2003}).

\bibitem[{\citenamefont{Stockman}(2004{\natexlab{a}})}]{Phys_Rev_Lett_93_2004_%
Tapered_Plasmonic_Waveguides}
\bibinfo{author}{\bibfnamefont{M.~I.} \bibnamefont{Stockman}},
  \bibinfo{journal}{Phys. Rev. Lett.} \textbf{\bibinfo{volume}{93}},
  \bibinfo{pages}{137404} (\bibinfo{year}{2004}{\natexlab{a}}).

\bibitem[{\citenamefont{Stockman}(2004{\natexlab{b}})}]{Stockman_Plasmonics_20%
04_Adiabatic_Concentration_SPIE}
\bibinfo{author}{\bibfnamefont{M.~I.} \bibnamefont{Stockman}}, in
  \emph{\bibinfo{booktitle}{Plasmonics: Metallic Nanostructures and Their
  Optical Properties II}}, edited by \bibinfo{editor}{\bibfnamefont{N.~J.}
  \bibnamefont{Halas}} \bibnamefont{and} \bibinfo{editor}{\bibfnamefont{T.~R.}
  \bibnamefont{Huser}} (\bibinfo{publisher}{SPIE}, \bibinfo{address}{Denver,
  Colorado}, \bibinfo{year}{2004}{\natexlab{b}}), vol. \bibinfo{volume}{5512},
  pp. \bibinfo{pages}{38--49}.

\bibitem[{\citenamefont{Babajanyan et~al.}(2000)\citenamefont{Babajanyan,
  Margaryan, and
  Nerkararyan}}]{Babajanyan_Margaryan_Nerkararyan_JAP_2000_SPP_on_Cone}
\bibinfo{author}{\bibfnamefont{A.~J.} \bibnamefont{Babajanyan}},
  \bibinfo{author}{\bibfnamefont{N.~L.} \bibnamefont{Margaryan}},
  \bibnamefont{and} \bibinfo{author}{\bibfnamefont{K.~V.}
  \bibnamefont{Nerkararyan}}, \bibinfo{journal}{J. Appl. Phys.}
  \textbf{\bibinfo{volume}{87}}, \bibinfo{pages}{3785} (\bibinfo{year}{2000}).

\bibitem[{\citenamefont{Maier et~al.}(2006)\citenamefont{Maier, Andrews,
  Martin-Moreno, and
  Garcia-Vidal}}]{Maier_Andrews_Martin-Moreno_Garcia-Vidal_PRL_2006_THz_SPPs_i%
n_Periodically_Corrugated_Wires}
\bibinfo{author}{\bibfnamefont{S.~A.} \bibnamefont{Maier}},
  \bibinfo{author}{\bibfnamefont{S.~R.} \bibnamefont{Andrews}},
  \bibinfo{author}{\bibfnamefont{L.}~\bibnamefont{Martin-Moreno}},
  \bibnamefont{and} \bibinfo{author}{\bibfnamefont{F.~J.}
  \bibnamefont{Garcia-Vidal}}, \bibinfo{journal}{Phys. Rev. Lett.}
  \textbf{\bibinfo{volume}{97}}, \bibinfo{pages}{176805}
  (\bibinfo{year}{2006}).

\bibitem[{\citenamefont{Gramotnev}(2005)}]{Gramotnev_JAP_2005_Adiabatic_SPP_Na%
nofocusing}
\bibinfo{author}{\bibfnamefont{D.~K.} \bibnamefont{Gramotnev}},
  \bibinfo{journal}{J. Appl. Phys.} \textbf{\bibinfo{volume}{98}},
  \bibinfo{pages}{104302} (\bibinfo{year}{2005}).

\bibitem[{\citenamefont{Verhagen et~al.}(2006)\citenamefont{Verhagen, Kuipers,
  and
  Polman}}]{Verhagen_Kuipers_Polman_Nano_Lett_2006_Tapered_Plasmonic_Waveguide}
\bibinfo{author}{\bibfnamefont{E.}~\bibnamefont{Verhagen}},
  \bibinfo{author}{\bibfnamefont{L.}~\bibnamefont{Kuipers}}, \bibnamefont{and}
  \bibinfo{author}{\bibfnamefont{A.}~\bibnamefont{Polman}},
  \bibinfo{journal}{Nano Lett.}  (\bibinfo{year}{2006}).

\bibitem[{\citenamefont{Johnson and Christy}(1972)}]{Johnson:1972_Silver}
\bibinfo{author}{\bibfnamefont{P.~B.} \bibnamefont{Johnson}} \bibnamefont{and}
  \bibinfo{author}{\bibfnamefont{R.~W.} \bibnamefont{Christy}},
  \bibinfo{journal}{Phys. Rev. B} \textbf{\bibinfo{volume}{6}},
  \bibinfo{pages}{4370} (\bibinfo{year}{1972}).

\bibitem[{\citenamefont{Landau and
  Lifshitz}(1984)}]{Landau_Lifshitz_Electrodynamics_Continuous:1984}
\bibinfo{author}{\bibfnamefont{L.~D.} \bibnamefont{Landau}} \bibnamefont{and}
  \bibinfo{author}{\bibfnamefont{E.~M.} \bibnamefont{Lifshitz}},
  \emph{\bibinfo{title}{Electrodynamics of Continuous Media}}
  (\bibinfo{publisher}{Pergamon}, \bibinfo{address}{Oxford and New York},
  \bibinfo{year}{1984}).

\bibitem[{\citenamefont{Lerosey et~al.}(2007)\citenamefont{Lerosey, de~Rosny,
  Tourin, and
  Fink}}]{Lerosey_de_Rosny_Tourin_Fink_Science_315_1120_2007_Microwave_Time_Re%
versal_Subwavelength_Focusing}
\bibinfo{author}{\bibfnamefont{G.}~\bibnamefont{Lerosey}},
  \bibinfo{author}{\bibfnamefont{J.}~\bibnamefont{de~Rosny}},
  \bibinfo{author}{\bibfnamefont{A.}~\bibnamefont{Tourin}}, \bibnamefont{and}
  \bibinfo{author}{\bibfnamefont{M.}~\bibnamefont{Fink}},
  \bibinfo{journal}{Science} \textbf{\bibinfo{volume}{315}},
  \bibinfo{pages}{1120} (\bibinfo{year}{2007}).

\end{thebibliography}
\end{document}